\documentstyle[aps,prl,psfig,epsfig
]{revtex}
\input{epsf}
\draft
\begin{document}
\title{Fano resonances with discrete breathers}
\author{S. Flach$^1$, A. E.
Miroshnichenko$^1$, V. Fleurov$^2$ and M. V. Fistul$^3$}
\address{
$^1$ Max-Planck-Institut f\"ur Physik komplexer Systeme, N\"othnitzer
Strasse 38, D-01187 Dresden, Germany \\
$^2$ Beverly and Raymond Sackler Faculty of Exact Sciences,
School of Physics and Astronomy, Tel Aviv University, Tel Aviv 69978, Israel \\
$^3$ Physikalisches Institut III,
Universit\"at Erlangen-N\"urnberg, D-91058, Erlangen, Germany }

\date{\today}
\wideabs{
\maketitle
\begin{abstract}
A theoretical study of linear wave  scattering by time-periodic
spatially localized excitations (discrete breathers (DB)) is
presented. We obtain that the wave propagation is strongly
influenced by a local coupling between an open and closed channels
generated by the DB. A peculiar effect of {\it total reflection}
occurs due to a Fano resonance when a localized state originating
from closed channels resonates with the open channel. For the
discrete nonlinear Schr\"odinger chain we provide with an
analytical result for the frequency dependence of the transmission
coefficient, including the possibility of resonant reflection. We
extend the analysis to chains of weakly coupled anharmonic
oscillators and discuss the relevance of the effect for electronic
transport spectroscopy of mesoscopic systems.
\end{abstract}
}


It is a well established fact that various {\em nonlinear
spatially discrete} systems can support time-periodic spatially
localized excitations called {\em discrete breather} states (DB)
\cite{reviews}. These states originate from a peculiar interplay
between the nonlinearity and discreteness of the lattice rather
than from a disorder. While the nonlinearity yields an
amplitude-dependent tunability of frequencies of DBs, $\Omega_b$,
the spatial discreteness of the system leads to finite upper
bounds for the frequency spectrum of small amplitude plane waves
$\omega_q$. This tunability allows one to escape resonances of all
multiples of the breather frequency $\Omega_b$ with the plane wave
frequencies $\omega_q$, and correspondingly to stabilize the DB
state. The frequency dependent localization length of DB's and
their stability with respect to small amplitude perturbations have
been widely studied \cite{reviews}. DBs have been observed in
experiments covering such diverse fields as interacting Josephson
junctions \cite{josephsons}, magnetic systems\cite{afm} and
lattice dynamics of crystals \cite{ptcl}.

For propagating linear waves a DB acts as a time-periodic
scattering potential, and the transmission coefficient $T$
depends on both the wave vector $q$ of the linear wave and the
breather frequency $\Omega_b$. The most peculiar effect, observed
in many numerical studies of wave scattering by DBs, is the {\em
total reflection} as $T~=~0$ \cite{scattering,scattering2}.
Note that the presence of a static potential cannot lead to such a
total reflection  in one-dimensional systems. Similar features are
also discussed in other areas, such as electron transport through
point contacts, quantum dots and wires \cite{gl93,ns94}. The
crucial condition allowing a total reflection in these systems is
the presence of a few coupled channels connected with the
transverse direction of motion. On the other hand, the wave
propagation in the presence of a time-periodic scattering
potential is characterized by {\em open and closed channels}
emerging from the Floquet formalism
\cite{reviews,scattering,scattering2}. The open channel guides the
propagating waves, while the eigenfrequencies of closed channels
do not match the spectrum of linear waves.

In this Letter we show that the total reflection of linear waves
in the open channel occurs when a {\it localized state}
originating from one of the closed channels resonates with the
open channel spectrum, a condition similar to the well-known Fano
resonance \cite{fano}. We will use this understanding to {\em
predict} Fano resonance positions for wave scattering by breathers
in weakly interacting anharmonic oscillator chains and to discuss
the relevance of this effect for electronic transport
spectroscopy.

We start our study of the wave scattering by DB with the discrete
nonlinear Schr\"odinger system (DNLS) that has been used
frequently to study breather properties due to its tractable form.
Wave scattering by breathers in the DNLS was studied numerically
in \cite{kimkim}, where resonant total reflection was also
observed. First, we provide an analytical solution for the DNLS
scattering problem. The equations of motion for the DNLS are
\begin{equation}
i\dot{\Psi}_n = C(\Psi_{n+1} + \Psi_{n-1}) + |\Psi_n|^2\Psi_n
\label{DNLS}
\end{equation}
where the integer $n$ labels the lattice sites, $\Psi_n$ is a
complex scalar variable and $C$ describes the nearest neighbor
interaction on the lattice. For small amplitude waves $\Psi_n(t) =
\epsilon e^{i(\omega_qt-qn)}$ the dispersion relation is
\begin{equation}
\omega_q = -2C \cos q~~.
\label{dispersion}
\end{equation}

Breather solutions have the form
\begin{equation}
\hat{\Psi}_n(t) = \hat{A}_n e^{-i\Omega_b t}\;,\;\hat{A}_{|n| \rightarrow
\infty} \rightarrow 0
\label{breather}
\end{equation}
where the time-independent amplitude $\hat{A}_n$ can be taken real
valued, and the breather frequency $\Omega_b \neq \omega_q$ is  a
function of the maximum amplitude $\hat{A}_0$. The spatial
localization is given by an exponential law $\hat{A}_n \sim
e^{-\lambda |n|}$ where $\cosh \lambda = |\Omega_b|/2C$
\cite{rsmsa94}. Thus the breather can be approximated as a
single-site excitation if $|\Omega_b| \gg C$. Then the relation
between the single-site amplitude $\hat{A}_0$ and $\Omega_b$
becomes $\Omega_b = \hat{A}_0^2$. In the following we will neglect
the breather amplitudes for $n \neq 0$, i.e. we assume
$\hat{A}_{n\neq 0} \approx 0$, since $\hat{A}_{\pm 1} \approx
\frac{C}{\Omega_b} \hat{A}_0\ll \hat{A}_0$. We emphasize that in
the limit of a nearly single-site localized breather solution in
the DNLS the solution {\it is} linearly stable \cite{rsmsa94}.

We add small perturbations to the breather solution
\begin{equation}
\Psi_n(t) = \hat{\Psi}_n(t) + \phi_n(t)
\label{linearization}
\end{equation}
and linearize equations (\ref{DNLS}) with respect to $\phi_n(t)$:
\begin{equation}
i\dot{\phi}_n = C(\phi_{n+1}+\phi_{n-1}) +\Omega_b \delta_{n,0}
(2\phi_0 + e^{-2i\Omega_b t} \phi_0^*)
\label{linequations}
\end{equation}
with $\delta_{n,m}$ being the usual Kronecker symbol. The general solution to
this problem is given by the sum of contributions due to two channels,
\begin{equation}
\phi_n(t) = X_n e^{i\omega t} + Y_n^* e^{-i(2\Omega_b +\omega)t}
\label{2channels}
\end{equation}
where $X_n$ and $Y_n$ are complex numbers satisfying the following algebraic
equations:
\begin{eqnarray}
-\omega X_n = C(X_{n+1}+X_{n-1}) +\Omega_b \delta_{n,0}
(2X_0 + Y_0)\;, \label{ae1} \\
(2\Omega_b + \omega) Y_n = C(Y_{n+1}+Y_{n-1}) +\Omega_b \delta_{n,0}
(2Y_0+X_0)\;. \label{ae2}
\end{eqnarray}

Away from the breather center $n=0$, Eq. (\ref{linequations})
allows for the existence of plane waves with the spectrum
$\omega_q$. Keeping in mind the propagation of waves, we set
$\omega \equiv \omega_q$ for some value of $q$. Thus the $X$
channel is an open one, while the $Y$ channel is a closed one,
i.e. its frequency $-(2\Omega_b+\omega_q)$ does not match the
spectrum $\omega_q$.

Instead of solving Eqs. (\ref{ae1}) and (\ref{ae2}) we will
consider a more general set of equations
\begin{eqnarray}
-\omega_q X_n = C(X_{n+1}+X_{n-1}) - \delta_{n,0}
(V_x X_0 + V_a Y_0)\;, \label{nae1} \\
(\Omega + \omega_q) Y_n = C(Y_{n+1}+Y_{n-1}) - \delta_{n,0}
(V_y Y_0+ V_a X_0) \label{nae2}
\end{eqnarray}
which is reduced to (\ref{ae1}) and (\ref{ae2}), if
$\Omega=2\Omega_b$ and $V_x=V_y=2V_a=-2\Omega_b$. For a particular
case $V_a=0$, i.e. when the closed $Y$-channel is decoupled from the
open one, the former possesses exactly one localized eigenstate
due to a nonzero value of $V_y$ with the frequency
\begin{equation}
\omega^{(y)}_L = -\Omega + \sqrt{V_y^2 + 4C^2}. \label{localmode}
\end{equation}

To compute the transmission coefficient $T$ we use the transfer matrix
method described, e.g. in Ref. \cite{bambihu}. The boundary
conditions are $ X_{N+1} = \tau e^{iq}\;,\;X_N = \tau\;,\;
Y_{N+1}=D/\kappa\;,\; Y_N = D $ for the right end of the chain and
$ X_{-N-1}=1+\rho \;,\; X_{-N} = e^{iq} + \rho e^{-iq}\;,\; Y_{-N-1} =
F\;,\; Y_{-N} = \kappa F $ for the left one. Here $\tau$ and $\rho$ are
the transmission and reflection amplitudes with $T=|\tau|^2=1-|\rho|^2$. 
$F$ and $D$ describe
the exponentially decaying amplitudes in the closed $Y$-channel,
where the degree of localization is described by the coefficient
$\kappa\equiv e^{\lambda}$:
\begin{equation}
\kappa = \frac{1}{2C}\left[ \Omega + \omega_q +
\sqrt{(\Omega+\omega_q)^2 - 4C^2} \right]\;.
\label{kappa}
\end{equation}
The $4\times 4$ transfer matrix is defined by Eqs.
(\ref{nae1}) and (\ref{nae2}) at $n=0$. After finding the
solutions of the corresponding four linear equations we obtain
\begin{eqnarray}
T = \frac{4 \sin ^2 q}{\left( 2\cos q - a -\frac{d^2
\kappa}{2-b\kappa} \right)^2 + 4\sin ^2 q}\;, \label{t1}
\\
a = \frac{-\omega_q + V_x}{C}\;,\;b=\frac{\Omega+\omega_q+ V_y}{C} \;,\; d
= \frac{V_a}{C}\;.
\label{t2}
\end{eqnarray}
This is the central result of this paper, which allows one to
conclude that total reflection is obtained when the condition
\begin{equation}
2-b\kappa=0
\label{zero1}
\end{equation}
is realized. It is equivalent to the resonance condition
\begin{equation}
\omega_q = \omega^{(y)}_L,
\label{resonance}
\end{equation}
which has a clear physical meaning: total reflection occurs
when a local mode, originating from the closed
$Y$-channel, resonates with the plane wave spectrum $\omega_q$ of
the open $X$-channel. The only condition is that the coupling
between the open and closed channels $V_a$ is nonzero. Remarkably
the resonance position does not depend on the actual value of
$V_a$, i.e. there is no renormalization. The existence of local
modes, which originate from the $X$-channel for nonzero $V_x$ and
possibly resonate with the closed $Y$-channel is evidently of no
importance. Eq.(\ref{t1}) also yields zero transmission for
$q=0,\pi$ due to a vanishing of the group velocity $d\omega_q/dq \sim
\sin q$
for these $q$ values; we will not focus on these trivial total reflections. 

This resonant total reflection is very similar to the Fano
resonance \cite{fano}, since it is directly related to a local
state resonating and interacting with the continuum of extended
states. The fact that the resonance location is independent of the
coupling $V_a$ is due to the local, single-site, character of the
coupling between the local mode (originating from the $Y$-channel)
and the open channel. If this coupling has a finite nonzero
localization length, i.e. several neighboring sites are involved,
then the total reflection happens at an energy, which does not
necessarily coincide with that of the local state
(\ref{resonance}) \cite{renormalization}. A more physical
formulation for the condition of absence of significant
renormalizations of the resonance location is that the wave length
of the propagating wave is large compared to the extension of the
space region where the channel coupling occurs.

Returning to the case of a DNLS breather at weak coupling, we
insert the values for $\Omega\;,\;V_x\;,V_y$ and $V_a$ into
(\ref{t1},\ref{t2}) and obtain the following expression for the
transmission coefficient:
\begin{equation}
T = \frac{4 \sin ^2 q}{\left( \frac{2\Omega_b}{C}
-\frac{\Omega_b^2}{2C^2}\frac{\kappa}{1+\kappa\cos q }
\right)^2 + 4\sin ^2 q}\;.
\label{tb1}
\end{equation}
The result is that any breather solution of the DNLS close to the
anticontinuous limit (the interaction $C$ goes to zero) provides
us with a total reflection in a close vicinity of $q=\pi /2$.  
Indeed, if we expand (\ref{tb1}) in
$\displaystyle\frac{C}{\Omega_b}$ we obtain
\begin{equation}
T \approx \frac{4C^4}{\Omega_b^4} \sin^2 2q
\label{tb2}
\end{equation}
in the lowest order, provided $\frac{C}{2\Omega_b} \ll | \cos q
|$.

\begin{figure}[htb]
\vspace{20pt}
\centerline{\psfig{figure=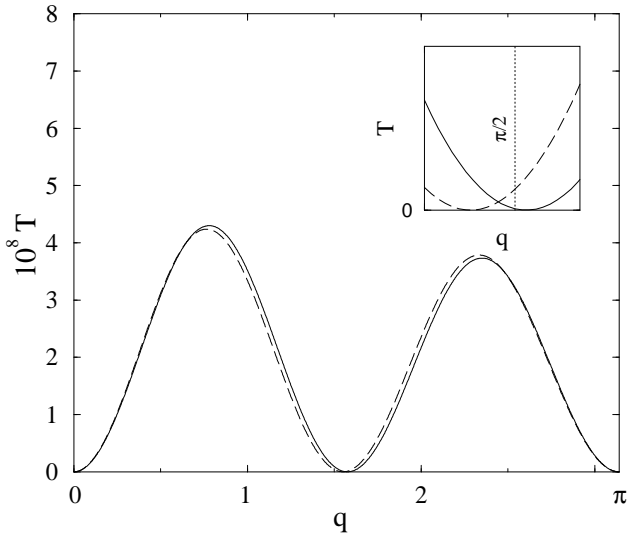,width=82mm,height=70mm}}
\vspace{10pt} \caption{Transmission coefficient $T$ for DNLS
breather with $C=0.01$ and $\Omega_b=1$ versus $q$. Dashed line -
numerically exact result, solid line - analytical result
(\ref{tb1}). Inset -  zoom around total reflection, where dotted
line marks $q=\pi/2$.} \label{fig1}
\end{figure}

Fig.\ref{fig1} compares a numerically obtained $q$ dependence of
the transmission coefficient with (\ref{tb1}) for $C=0.01$ \cite{fmf02}.
We obtain a very good agreement, except for a small shift of the true
total reflection position with respect to $q=\pi/2$.
It is due to the small but nonzero {\em finite extension} of the
scattering potential,
which leads to a spread of the coupling between the local mode
and the open channel.

Next we consider the system of weakly interacting anharmonic
oscillators:
\begin{equation}
\ddot{X}_n = -V'(X_n) + C(X_{n-1}+X_{n+1} - 2X_n)
\label{kg}
\end{equation}
where the oscillator potential $V$ possesses one minimum,
$V'(0)=0$ and $V''(0)=1$. The spectrum of small amplitude plane
waves is given by $\omega^2_q = 1+4C \sin^2 (q/2)$ and discrete
breather solutions are time-periodic spatially localized solutions
of Eq. (\ref{kg}) with frequency $\Omega_b \neq \omega_q/m$ for
any nonzero integer $m$.  Here we will again consider a weak
interaction between the oscillators ($C~<<~1$), and assume that
the breather is essentially a single-site excitation $\hat{X}_0(t)
= \hat{X}_0(t + 2\pi/\Omega_b) \neq 0$ and $\hat{X}_{n \neq 0}(t)
\approx 0$. The equations for the linearized phase space flow
around the breather solution are then given by \cite{scattering2}
\begin{eqnarray}
\ddot{\epsilon}_n = 
-\epsilon_n + C(\epsilon_{n-1}+ \epsilon_{n+1}
-2\epsilon_n) \nonumber \\
- \delta_{n,0} \left( V''(\hat{X}_0(t)) -1\right)
\epsilon_0\;. \label{linearkg}
\end{eqnarray}
For $C=0$ we expand $V''(\hat{X}_0(t)) = \sum_k v_k
e^{ik\Omega_bt}$ and use the Floquet representation $\epsilon_0(t)
= \sum_k e_{0k} e^{i(\omega+k\Omega_b)t}$ to arrive at the set of
equations,
\begin{equation}
-(\omega+k\Omega_b)^2 e_{0k} = -\sum_{k'} v_{k-k'}e_{0k'},
\label{kgset}
\end{equation}
for the site $n=0$, at which the breather is excited. This
complete set of linear equations describes the coupling of the
open channel (the corresponding amplitude of oscillations
$e_{00}$) and many closed channels (the amplitudes of oscillations
$e_{0k}$ with $k \neq 0$).

To obtain the condition for total reflection we apply a procedure,
which is similar to our analysis of the DNLS system. We "turn off" the
coupling between the open channel and closed channels, i. e. the
amplitude $e_{00}$ is set to zero in Eq. (\ref{kgset}). The
remaining homogeneous set of equations determines the local
states $\omega_L$ of closed channels. The values of
$\omega=\omega_L$ are those for which the corresponding
determinant vanishes. Moreover, we are interested in the situation
when $\omega_L^2=1$, which results in a resonance of a local mode
of the system of closed channels with the open channel $\omega_q^2
\approx 1$. It is evident that such a situation is not necessarily
realized for an arbitrary value of $\Omega_b$. To find the proper
value of $\Omega_b$ we put $\omega=1$ and scan $\Omega_b$ for a
given potential $V$ (note that the parameters $v_k$ depend on
$\Omega_b$). As a result, we expect to find a discrete set of
$\Omega_b$ values, for which the determinant of the reduced set of
equations Eq. (\ref{kgset}) zeros, and a Fano resonance occurs.

Thus at variance to the DNLS case, a Klein-Gordon chain in the
anticontinous limit will provide us with a total reflection of
waves by breathers only for a selected discrete set of breather
frequencies. Increasing the coupling $C$ transforms each of these
frequency values into frequency stripes on the real axis, which
will continue to increase with $C$. This follows from the fact
that the band width of $\omega_q$ increases linearly with the
coupling $C$, whereas the shift of the eigenvalues is proportional
to $C^2$. 

Note here, that the Green function technique elaborated
for the wave scattering by DBs in Ref. \cite{scattering2}, leads
to the same procedure described above in the limit of small $v_k$,
allowing to obtain the dependence of $\omega_L$ on the amplitude
and frequency of the DB.

\begin{figure}[htb]
\vspace{20pt}
\centerline{\psfig{figure=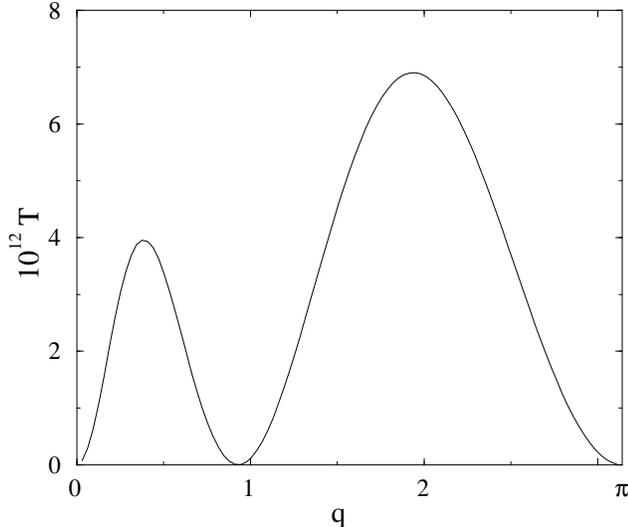,width=82mm,height=70mm}}
\vspace{2pt} \caption{Transmission coefficient versus $q$
for a Klein-Gordon chain breather with $\Omega_b=1.38$ and
$C=0.001$. } \label{fig2}
\end{figure}

In order to demonstrate the validity of our constructive approach,
we choose
\begin{equation}
V(x)=\frac{1}{2}x^2 + \frac{1}{3}x^3 + \frac{1}{4}x^4
\label{potential}
\end{equation}
and carry out the above procedure. We obtain $\Omega_b \approx
1.38$. The prediction then is that for small $C$ a breather with
$\Omega_b$ close to this value yields a total reflection. The
numerical result for the transmission is shown in Fig \ref{fig2}
for $C=0.001$ \cite{fmf02}. We indeed observe a total reflection
around $q\approx 1$, as predicted.

Our analysis can be directly used for the detection of Fano resonances
in the propagation of linear
waves in weakly dissipative Josephson junction arrays in the presence
of discrete breathers \cite{josephsons},\cite{scattering2}. 
Moreover our results
allow to formulate optimal conditions for
electronic transport spectroscopy of molecules, quantum dots or
optical cavities, coupled to a system of leads. Indeed a one
dimensional electronic transport through the localized states in
the presence of externally applied microwave radiation of
frequency $\Omega$, will display a Fano resonance (and
correspondingly zero conductivity) as the condition $E_F=\Omega
\pm E_n$ is satisfied. Here, $E_F$ is the Fermi energy, and $E_n$
are the localized levels. To establish this type of a spectroscopy
leads should be quasi-one-dimensional such that only one
transversal mode persists. Then we reduce the lead system to one
open channel. The other important condition is to satisfy that the
wavelength of excitations (e.g. at the Fermi energy for electronic
transport) is {\em large} compared to the spatial extension of the
coupling to the localized states. Additional optional spatial
modulations, e.g. in the leads, may be used to generate artificial
gaps in the electronic spectrum and thus to tune the wavelength.
The power of microwave radiation has to be rather small to exclude
the renormalization of $E_n$. The coupling between propagating
states and localized states may be also provided by, e.g.
spin-spin or spin-orbit interactions. In this case zero
conductance will be the consequence of a bare dot state passing
the Fermi energy ($E_F=E_n$).
\\
\\
\\
Acknowledgements
\\
We thank K. Kikoin, I. Rotter and M. Titov for useful discussions.


\begin{references}
\bibitem{renormalization}
To find this result we may look at the derivation of the
renormalization of the Fano resonance in Ref.\cite{ns94} for the
case of electron scattering in quantum wires. The shift of the
resonance is given there by the integral in the last line of
Eq.(A11), which zeros if the coupling potential is $\delta$-like.
This analysis can be redone for discrete systems, which we
consider here, and the same zero shift will be obtained for the
single-site coupling (Kronecker symbol appears instead of the
$\delta$-function). However, if the coupling involves several
sites a finite shift of the total reflection with respect to the
local mode due to the closed channel is obtained.
\bibitem{fmf02}
For details concerning the numerical methods of obtaining high
precision transmission data we refer to \cite{scattering2}.
\end{references}
\end{document}